\documentstyle[10pt,epsfig,supertabular]{elsart}
\input{epsf.sty}
\newcommand{\lu}{{\it L}}

\newcommand{\Laena}{\Lambda_{\mathrm Ae}N_{\mathrm A}}

\newcommand{\Lahna}{\Lambda_{\mathrm Ah}\Delta N_{\mathrm A}}

\newcommand{\Ldendnn}{\Lambda_{\mathrm De}\left(N_{\mathrm
D}+N_{\mathrm n}\right)}

\newcommand{\Ldhndnn}{\Lambda_{\mathrm Dh}\left(N_{\mathrm
D}+N_{\mathrm n}\right)}
\newcommand{\Ldnd}{\Lambda_{\mathrm D}N_{\mathrm D}}

\newcommand{\LRnh}{\Lambda_{\mathrm R}n_{\mathrm h}\left(x,t\right)}

\newcommand{\LRne}{\Lambda_{\mathrm R}n_{\mathrm e}\left(x,t\right)}

\begin{document}
\begin{frontmatter}
\title {Response of CsI(Tl) scintillators over a large range in energy
and atomic number of ions (Part I): recombination and $\delta$ -- electrons}
\small
\author[buch,ipno]{M.~P\^arlog},
\author[ipno]{B.~Borderie\thanksref{cores}}, 
\author[ipno]{M.F.~Rivet},
\author[buch,ipno]{G.~T\u{a}b\u{a}caru},  
\author[ganil]{A.~Chbihi},
\author[maroc]{M. Elouardi},
\author[lpc]{N.~Le Neindre},
\author[lpc]{O.~Lopez}, 
\author[ipno]{E.~Plagnol}, 
\author[ipno]{L.~Tassan-Got},
\author[ganil]{G.~Auger},
\author[ipno]{Ch.O.~Bacri}, 
\author[lpc]{N.~Bellaize}, 
\author[lpc]{F.~Bocage},
\author[lpc]{R.~Bougault}, 
\author[ganil]{B.~Bouriquet}, 
\author[lpc]{R.~Brou}, 
\author[cea]{P.~Buchet},
\author[cea]{J.L.~Charvet}, 
\author[lpc]{J.~Colin}, 
\author[lpc]{D.~Cussol}, 
\author[cea]{R.~Dayras},
\author[ipnl]{A.~Demeyer}, 
\author[cea]{D.~Dor\'e}, 
\author[lpc]{D.~Durand}, 
\author[ganil]{J.D.~Frankland},
\author[ipno,cnam]{E.~Galichet}, 
\author[lpc]{E.~Genouin-Duhamel}, 
\author[ipnl]{E.~Gerlic},
\author[ganil]{S.~Hudan}, 
\author[ipnl]{D.~Guinet}, 
\author[ipnl]{P.~Lautesse}, 
\author[ipno]{F.~Lavaud},
\author[ganil]{J.L.~Laville}, 
\author[lpc]{J.F.~Lecolley}, 
\author[ipnl]{C.~Leduc}, 
\author[cea]{R.~Legrain},
\author[lpc]{M.~Louvel}, 
\author[ipnl]{A.M.~Maskay}, 
\author[cea]{L.~Nalpas}, 
\author[lpc]{J.~Normand},
\author[lpc]{J.~P\'{e}ter}, 
\author[napol]{E.~Rosato}, 
\author[ganil]{F.~Saint-Laurent\thanksref{a}},
\author[lpc]{J.C.~Steckmeyer}, 
\author[lpc]{B.~Tamain}, 
\author[ganil]{O.~Tirel},
\author[lpc]{E.~Vient}, 
\author[cea]{C.~Volant}, 
\author[ganil]{J.P.~Wieleczko}
\collaboration{(INDRA collaboration)}
\address[buch]{National Institute for Physics and Nuclear Engineering,
RO-76900 Bucharest-M\u{a}gurele, Romania}
\address[ipno]{Institut de Physique Nucl\'eaire, IN2P3-CNRS, F-91406 Orsay
 Cedex, France.}
\address[ganil]{GANIL, CEA et IN2P3-CNRS, B.P.~5027, F-14076 Caen Cedex, France.}
\address[maroc]{Laboratoire de Physique Nucl\'eaire Appliqu\'ee,
Kenitra, Maroc.}
\address[lpc]{LPC, IN2P3-CNRS, ISMRA et Universit\'e, F-14050 Caen Cedex,
France.}
\address[cea]{DAPNIA/SPhN, CEA/Saclay, F-91191 Gif sur Yvette Cedex,
France.}
\address[ipnl]{Institut de Physique Nucl\'eaire, IN2P3-CNRS et Universit\'e,
F-69622 Villeurbanne
Cedex, France.}
\address[napol]{Dipartimento di Scienze Fisiche e Sezione INFN, Universit\`a
di Napoli ``Federico II'', I80126 Napoli, Italy.}
\address[cnam]{Conservatoire National des Arts et M\'etiers, F-75141
Paris cedex 03.}
\thanks[cores]{Corresponding author. Tel 33 1 69157148; fax 33 1
69154507; e-mail borderie@ ipno.in2p3.fr} 
\thanks[a]{present address: DRFC/STEP, CEA/Cadarache, F-13018
Saint-Paul-lez-Durance Cedex, France.}
\begin{abstract}
A simple formalism describing the light response of CsI(Tl) to heavy ions, 
which quantifies the luminescence and the quenching in 
terms of the competition between radiative transitions following the carrier 
trapping at the Tl activator sites and the electron-hole recombination,  
is proposed. The effect of the $\delta$ rays on the scintillation 
efficiency is for the first time quantitatively included in a fully consistent 
way. The light output expression depends on four parameters determined by a
procedure of global fit to experimental data.
\end{abstract}
\end{frontmatter}
\section{Introduction\label{sect1}}
The thallium-activated caesium iodide (CsI(Tl)) scintillation crystals combine 
the advantages of high stopping power and high reliability for large thickness, 
 good energy resolution (a few percent) and the possibility of light particle 
isotopic identification using the pulse shape discrimination technique. The 
response of these detectors has a non-linear dependence on the energy of the 
incident particle. The reduction of the light yield with regard to the
linear behaviour is referred to as ``quenching''. In addition, for a given 
energy, the light output depends on the type of the particle.
\par
Obtaining a particle dependent light output description and energy 
calibration of these scintillators, over a very large range of energies and 
ion atomic numbers, is the primary objective of the present work. 
Quite successful early approaches of the main experimental 
observations to be considered are discussed in section \ref{sect6}.
A light response expression depending on the mass, charge and energy of the
ionizing particle is derived in section \ref{sect3}, in a simple formalism 
which makes use of band theory scheme in insulators. The quenching is
connected to the carrier recombination or their trapping at crystal 
imperfection sites, eventually enhanced by the nuclear interaction of the 
slowing down fragment with lattice nuclei. The contribution of the generated 
$\delta$ -- rays is quantitatively included. The capability of this 
``recombination and nuclear quenching model'' (RNQM) to describe experimental 
data is demonstrated in section \ref{sect4}. These data were obtained by means 
of the INDRA array, through a procedure described in the accompanying paper 
\cite{par00ii}. A summary is given in section \ref{sect5}.
\newpage							
Notation and values of physical constants and variables used in this paper.
{\small
\tablefirsthead{
 \hline
Symbol$^*$ & Definition & Units or Value \\
 \hline}
\tablehead{
 \hline
Symbol$^*$ & Definition & Units or Value \\
 \hline
 }
\tabletail{
 \hline
 }
 \par
\begin{supertabular}{cp{7cm}p{4cm}}
\multicolumn{3}{c}{\it Physical constants} \\
$c$ & speed of light in vacuum &  $299~792~458$ m~s$^{-1}$  \\
$u$ & unified atomic mass unit &  $931.494~32$~MeV/$c^2$ \\
$m_e$ & electron mass & $0.510~999~06$ MeV/$c^2$ \\
$\cal N_A$ & Avogadro constant & $6.022~136~7\times$10$^{23}$~mol$^{-1}$ \\
$k$ & Boltzmann constant & $8.617~385\times$10$^{-5}$~eV~K$^{-1}$ \\
$h$ & Planck constant & $41.356~692\times$10$^{-22}$~MeV~s \\
\multicolumn{3}{c}{\it CsI(Tl) crystal} \\
$\rho$ & CsI crystal density & 4.53 g~cm$^{-3}$ \\ 
$<A_{CsI}>$ & average atomic mass number for CsI & 129.905  \\
$<Z_{CsI}>$ & average atomic number & 54 \\
$N_H$ & host cation and anion volume concentrations & ions~cm$^{-3}$  \\
$N_{A0}$ & Thallium activator volume concentration & ions~cm$^{-3}$  \\
$C$ & Thallium molar concentration & \% \\
$\epsilon$ &  energy gap between the valence and  
conduction bands & $\approx$ 10 eV \cite{mey61} \\
\multicolumn{3}{c}{\it Charged particle passing through the CsI(Tl) crystal} \\
$A$ & atomic mass number  &  \\
$Z$ & atomic number  &	\\
$E$ & current kinetic energy in the CsI(Tl) crystal & MeV \\
$x$ & coordinate along the path & 10$^{15}$ atoms~cm$^{-2}$ \\
$E_0$ &  experimental initial kinetic energy in the 
crystal & MeV \\
$R(E_0)$ & range & 10$^{15}$ atoms~cm$^{-2}$ \\
$Q_0$ & approximate total integrated charge $\propto$ experimental light output & a.u. \\
$\alpha,~\beta$ & relativistic kinematic variables & \\
$e_{\delta}$ & energy per nucleon threshold for $\delta$ - ray production & 
MeV/$u$ \\
$E_{\delta}$ & energy threshold for $\delta$ - ray production & MeV\\
$\alpha_{\delta}, \beta_{\delta}$ &  the corresponding relativistic kinematic 
variables & \\					 
$\Delta E$ & energy deposited in the preceding  
detection layer & MeV \\
\multicolumn{3}{c}{\it $\delta$ -- rays} \\
$T$ & kinetic energy transferred to an electron & \\
$T_{max}$ & maximum kinetic energy transferable to an 
electron & \\
$T_{cut}$ &  value of $T$ above which an electron 
is considered as a $\delta$ - ray & \\
$R_e$ & electron range & mg~cm$^{-2}$  \\
$r_c$ & radius of the highly ionized primary column & cm \\ 
\multicolumn{3}{c}{\it Stopping powers} \\
$\Delta$ & density effect in Bethe-Bloch formula & \\
$Z_{eff}$ &  effective charge of the impinging particle & \\
$\gamma_{eff}$ & coefficient of effective charge & \\
$I$ & ionization constant for CsI & 579.8 eV  \\
$-\d E/\d x$ & specific energy loss & eV$\times 10^{-15}$~atoms$^{-1}$~cm$^2$  \\
$-(\d E/\d x)_e = S_e$ & specific electronic stopping power & eV$\times
10^{-15}$~atoms$^{-1}$~cm$^2$  \\
$-(\d E/\d x)_n = S_n$ & specific nuclear stopping power & eV$\times
10^{-15}$~atoms$^{-1}$~cm$^2$  \\
$\rho S_{e,n}$ & electronic or nuclear stopping power & MeV~cm$^{-1}$ \\
$-(\d E/\d x)_{\mathrm T<T_{cut}}$ & restricted specific electronic stopping power &
eV$\times 10^{-15}$~atoms$^{-1}$~cm$^2$  \\
$-(\d E/\d x)_p$ &  part of the specific electronic energy 
loss remaining inside the primary column & eV$\times 10^{-15}$~atoms$^{-1}$~cm$^2$  \\
$-(\d E/\d x)_\delta $ & part of the specific electronic energy loss carried
outside the primary column by the $\delta$ -- rays & eV$\times
10^{-15}$~atoms$^{-1}$~cm$^2$ \\
${\cal F}$ & fractional energy loss transferred to a $\delta$ -- ray & \\
$\epsilon_{\mathrm n}$ & mean energy required to produce a dislocation & eV \\
$N_n$ & concentration of anion or cation defects produced by nuclear 
interaction & ions~cm$^{-3}$  \\
$\theta^\prime$ & c.m. deflection angle & rad \\
$\d \sigma^\prime/\d \Omega^\prime$ & c.m. Rutherford cross section & cm$^{-2}$ \\
$N_{Ruth}$ & concentration of anion or cation defects produced by 
Coulomb scattering & ions~cm$^{-3}$  \\ 
\multicolumn{3}{c}{\it Calculated light output and related variables} \\
$L$ & calculated integral light output & a.u. \\
$\cal S$ & scintillation efficiency factor of Birks & a.u. \\
$\cal KB$ & quenching factor of Birks & MeV$^{-1}$~cm$^2$~g$^{-1}$  \\
$(\d L/\d x)_p$ & infinitesimal light output inside the primary column & a.u. \\
$(\d L/\d x)_{\delta}$ & infinitesimal light output outside the primary column & a.u. \\
$h\nu$ & scintillation energy & eV \\
$W_D$ & energy of a nearest-neighbour bound in solid & $\approx$ 1 eV \\
$N_D$ & concentration of anion or cation defects produced by thermal 
vibrations & ions~cm$^{-3}$  \\
$\cal T$ & absolute temperature & $^{\circ}$K \\
$t$ & time & s \\
$n_e$ & current volume concentration of free electrons created by
the ionizing particle & electrons~cm$^{-3}$ \\
$n_h$ & current volume concentration of holes created by
the ionizing particle & holes~cm$^{-3}$ \\
$n_0=n_{e0}=n_{h0}$ & initial volume concentration of carriers created by the ionizing
particle & electrons~cm$^{-3}$ \\
$N_A$ & current volume concentration of electron traps at Tl activator sites &
ions~cm$^{-3}$  \\
$\Delta N_A$ & current volume concentration of hole traps at Tl activator sites &
ions~cm$^{-3}$  \\
$\Lambda_{Ae}$ & coefficient of electron trapping probability at activator
sites & s$^{-1}$~cm$^3$ \\
$\Lambda_{Ah}$ & coefficient of hole trapping probability at activator sites &
s$^{-1}$~cm$^3$ \\
$\Lambda_{De}$ & coefficient of electron trapping probability at defect sites &
s$^{-1}$~cm$^3$ \\
$\Lambda_{Dh}$ & coefficient of hole trapping probability at defect sites &
s$^{-1}$~cm$^3$ \\
$\Lambda_D$ & coefficient of trapping probability at defect
sites for both type of carriers & s$^{-1}$~cm$^3$ \\
$\Lambda_R$ & coefficient of carrier recombination probability &
s$^{-1}$~cm$^3$ \\
$\Lambda^\prime_{\mathrm Ae,D,R}$ & $\Lambda_{\mathrm
Ae,D,R}/(\Laena+\Ldnd)$ & cm$^3$ \\
$E_{calc}$ & calculated initial kinetic energy in the
CsI(Tl) crystal & MeV \\
$a$ & multiplicative constant & a.u. \\
$a_G$ & gain fit parameter in the exact expression of L & a.u. \\
$a_R$ & recombination quenching fit parameter in the exact expression of L &
eV$^{-1}10^{15}$atoms$^{-1}$cm$^2$ \\
$a_n$ & nuclear quenching fit parameter in the exact expression of L &
eV$^{-1}10^{15}$atoms$^{-1}$cm$^2$ \\
\hline
\end{supertabular}\\
$^*$Most of the notations of the original references have been kept.
}
\section{The total light response of CsI(Tl): two historical 
hints\label{sect6}}
\subsection{Quenching\label{subsect61}}
The CsI(Tl) light response to an ionizing charged particle of atomic number
$Z$ and mass number $A$ is non-linear with respect to the incident energy
$E_0$, depending on $Z$ and $A$ \cite{hal57,bas58,qui59}. The quenching has
been associated with the specific energy loss of the particle. This property
provides a means for the pulse shape discrimination technique. The
differential light output per unit path length expression
$\d L/\d x=\cal S|\d E/\d x|_e/(1 + \cal KB|\d E/\d x|_e)$, proposed by Birks 
for alpha particles in organic scintillators \cite{bir60} and applied 
also, without rigorous support, to inorganic ones \cite{bir64},
leads to the differential scintillation efficiency:
\begin{equation}
\frac{\d \lu}{\d E} ={\cal S}\frac{1}{1+{\cal KB}|\d E/\d x|_e} ,
\label{ecu6}
\end{equation}
where $\cal S$ and ${\cal KB}$ are the scintillation efficiency and the quenching
factors, respectively \cite{bir64}. The differential scintillation efficiency
is monotonically decreasing with specific electronic stopping power
$-(\d E/\d x)_{\mathrm e}$. This relation was applied with reasonable results 
in case of light  charged particles or intermediate mass fragments 
($Z \leq 15$) \cite{bir64,edf94}, i.e. as long as the fraction of the light 
yield from $\delta$-rays is not significant. Under the approximation 
$|\d E/\d x|_e \propto AZ^2/E$, Eq. (\ref{ecu6}) was integrated and 
successfully used to describe the total light output from light ions 
\cite{hor92}.
\subsection{Knock-on electrons\label{subsect62}}
Experimental studies, performed with ions of a few MeV/nucleon, as heavy as 
those from Boron to Neon \cite{new60,new61}, have shown that the scintillation 
efficiency of alkali halide crystals (NaI(Tl), CsI(Tl)) is not a function of 
$-(\d E/\d x)_{\mathrm e}$ alone,  but instead, is composed of a series of 
discrete functions, one for each incident particle. The observation has been 
thoroughly analyzed by Murray and Meyer \cite{mey62}, as shown in this 
subsection.   
\par
\begin{figure}
\epsfxsize=12.cm
\epsfysize=12.cm
\epsfbox{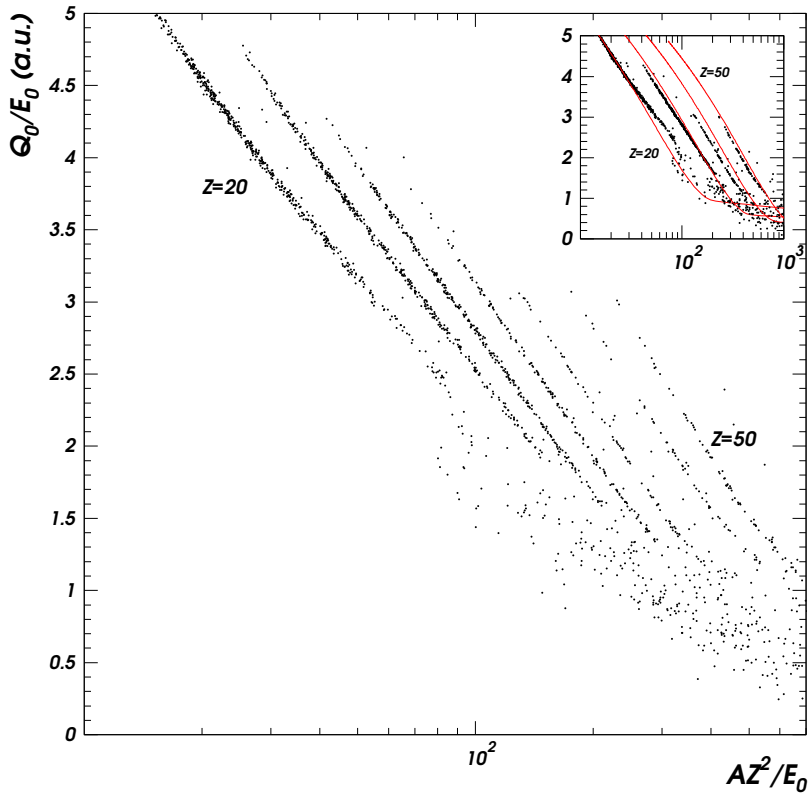}
\caption{Experimental light efficiency versus the estimate of the average
	specific electronic stopping power, for different fragments 
	($\Delta Z= 5$). Data from Xe + Sn system at 32 and 50 AMeV 
	\cite{par00ii}. The discrete loci are the effect of the 
	$\delta$ -- rays. Its description in the framework of the present 
	scintillation model - solid lines - is compared to the experimental 
	data in the upper box ($\Delta Z= 10$).}
\label{fig3}
\end{figure}

This effect is illustrated with INDRA data from Ref. \cite{par00ii} in
Fig. \ref{fig3}, where the experimental light efficiency ($Q_0/E_0$) is shown,
for different heavy ions, versus the estimate of  the average specific
electronic stopping power $AZ^2/E_0$ \cite{par00ii}.
$Q_0$ is the experimental light response corresponding to $E_0$. 
There is a domain where, at the same average specific energy loss, the 
higher the atomic number $Z$, the higher the scintillation efficiency. 
The above mentioned discrete functions have been interpreted
as a signature of modestly energetic ($\approx$ 1 keV) knock-on electrons,
the alternative term for $\delta$ - rays, whose ranges exceed 
the radius $r_c$ of the high carrier concentration column along the wake of
the ionizing particle, estimated to be around 40~nm~\cite{mey62}.
The differential light output per unit path length is the sum of two
contributions: one from the highly ionized primary column and one
from the knock-on electrons. Considering 
the total specific electronic stopping power of the incident particle:
$(-\d E/\d x)_e=(-\d E/\d x)_p+(-\d E/\d x)_\delta$ as the sum of the energy 
deposited inside the primary column $(-\d E/\d x)_p$ and the energy carried 
off by the $\delta$ - rays $(-\d E/\d x)_\delta$, 
the fractional energy loss $\cal F$ of the incident particle, 
deposited outside the primary column, was defined \cite{mey62}:
\begin{equation}
{\cal F}=\frac{(-\d E/\d x)_\delta}{(-\d E/\d x)_e} .
\label{ecu7}
\end{equation}
The differential scintillation efficiency is then given by:
\begin{equation}
\frac{\d \lu}{\d E} =(1-{\cal F})\left(\frac{\d \lu}{\d E}\right)_p + 
{\cal F}\left(\frac{\d \lu}{\d E}\right)_\delta .
\label{ecu8}
\end{equation}
The scintillation efficiency referring to the primary column $(\d \lu/\d E)_p$
may be treated with formula (\ref{ecu6}) provided that $(-\d E/\d x)_p$ is
known. The $\delta$ -- ray scintillation efficiency $(\d \lu/\d E)_\delta$ is
very nearly a constant for energies higher than 1 keV \cite{mey61,zer61}.
Estimations of $\cal F$ 
have shown that this quantity depends only on the energy per nucleon $E/A$ of 
the particle and it is different from zero only above a certain threshold value
$e_\delta=(E/A)_{\mathrm threshold}$ \cite{mey62}. These characteristics will
be found again in the next section in a very simple approach for calculation of
$\cal F$, valid for all ions.
\par
Thus, for a particle slowing down in a CsI(Tl) detector from the
incident velocity $\propto \sqrt{E_0/A} > \sqrt{e_\delta}$ to zero, the
differential scintillation efficiency is given by Eq. (\ref{ecu8}) as
long as $(E/A) > e_\delta$, and by Eq. (\ref{ecu6}) in the last part of
its range, when $(E/A) \leq e_\delta$. The total light
output would be obtained by integrating the scintillation efficiency
over the entire range of the particle. One has to note that the above formulae 
(\ref{ecu6}), (\ref{ecu8}) are not time dependent
and consequently they cannot predict the time behaviour of a scintillator
signal. In the next section, we shall develop a simple time dependent
formalism leading to scintillation efficiency expressions which, up to a
supplementary nuclear quenching term, reduces in the first order
approximation, to relations very similar to equations (\ref{ecu6}),
(\ref{ecu8}), when integrated over time.
\section{Recombination and nuclear quenching model\label{sect3}}
\subsection{Experimental evidence\label{subsect31}}
The 
CsI scintillator is an alkali halide crystal having a body centered cubic 
structure. The present work concerns data from CsI(Tl) operated at room 
temperature and having a given molar concentration 
$C\approx 0.02\% - 0.2\%$ \cite{bdh}. 
The corresponding thallium volume concentration is
$N_{A0}\approx 2.11\times 10^{18} - 2.10\times 10^{19}$ ions~cm$^{-3}$. 
 Experimental observations 
related to the scintillation and its behaviour as a function of temperature in 
both pure CsI and activated CsI(Tl) crystals can be found in Refs.
\cite{bir64,rob61ii,gwi63ii,mor60,bes62,bon52,kno57,kit66,sch90,val93,gwi63i,eby54}.
In the latter case, the thallium concentration dependence is regarded too.
At room temperature, the emission spectrum of a CsI(Tl) excited by 
charged projectiles \cite{gwi63ii} is dominated by a broad yellow band, 
centered near 550~nm.~ This band is thus associated with emission from Tl 
luminescence centres and is referred to as the Tl band \cite{gwi63ii}. 
The yellow band nearly completely replaces the competing blue band, related 
to $I^-$ vacancies or elementary colour centres \cite{sch90,val93} and the 
ultraviolet (UV) band, considered as an intrinsic property of a perfect crystal 
\cite{gwi63ii}. Both of the latter bands are present in case of pure CsI.
\par
We briefly remind the experimental evidence related to CsI(Tl) 
at a given temperature and concentration, on which our model is dependent.
 a) The scintillation efficiency is non-linearly increasing with the increasing 
energy of the particle. For different ions at the same energy, the 
scintillation efficiency 
decreases with increasing specific electronic stopping power~\cite{bir64}. If 
the ionizing particle is an electron or a $\gamma$ -- ray of energy higher 
than 100 keV, the light output may be considered to a good approximation as a 
linear function of energy~\cite{gwi63i,roo97}. b) The scintillation efficiency 
depends on the fraction of deposited energy creating $\delta$ -- rays 
\cite{mey62}, which affects the monotonic behaviour resulting from the
previous item. c) Combined studies of light efficiency and scintillator
decay-constant versus temperature \cite{bir64,sci58} have indicated that the
quenching effect is external to the luminescence centre \cite{bir64}.
\subsection{The scenario\label{subsect32}}
An ionizing particle excites the scintillator crystal along its path by 
creating carriers: electrons, promoted in the conduction band, and associated 
holes in the valence band. CsI(Tl) is an insulator and the energy gap between 
these two bands is of the order of the mean ionization energy 
$\epsilon\approx 10$ eV \cite{mey61}. Thallium iodide is melted with the 
caesium iodide host bulk before the crystal growth \cite{bdh}. Some cation 
sites are occupied by Tl$^+$ ions having lost one 
valence electron in favour of I$^-$ anions. As in the case of KCl(Tl) 
\cite{joh53}, the excited energy levels of the Tl$^+$ ions are presumably 
localized in the forbidden band of the host lattice. Even though deeper than 
in the semiconductor case, these levels may essentially participate in the 
ionized lattice deexcitation, provided that they are free.    
\par
Although the emission bands of the CsI(Tl) scintillators, of which the most
prevalent one is centred at about 550~nm,~ have been intensively studied,
there is still controversial discussion about the origin of this band 
\cite{joh53,pop97,mas68,der99,gut74,kau70,mur75,mic94,mic95,smo79,spa94,nag94,nag95}. 
Thus, according to different interpretations, the luminescence emission may 
imply, for example, a single thallium centre \cite{der99}, a couple of 
Tl$^+$ - $V_k$ centres \cite{spa94} or even more complex cluster configurations 
nearby a Tl impurity \cite{nag94,nag95}. As definitions of the imperfection 
centres we use those in Ref. \cite{mar79}.
Most of the mentioned approaches converge towards two ideas: for the
luminescence induced by highly ionizing particles, both type of carriers,
electrons and holes, play a role in the emitted 
light, i.e. in the scintillation efficiency; the radiative emission around 
550~nm~ takes place in the $(Tl^+)^*$ excited ions. Let us imagine a
formal, oversimplified scenario for the scintillation process, by
exemplifying it on the latter luminescence centres.
\par
We suppose that the Tl$^+$ ion, with its 2 remaining valence electrons,
 superfluous in the ionic bond of the CsI lattice, has a donor behaviour.     
Stimulated by thermal vibrations, the Tl$^+$ ions are loosing, in part,
 some of the outer electrons, creating intruded levels into the upper part of 
the forbidden band. These double ionized thallium atoms Tl$^{++}$ would become 
electron traps of volume concentration $N_A$. The remaining Tl$^+$ ions would 
be the hole traps, with the volume concentration $\Delta N_A=N_{A0}-N_A$. One 
could reason as following: in the highly ionized fiducial volume, 
the holes are mainly trapped, via a mechanism which is disregarded here, by 
Tl$^+$ ions becoming Tl$^{++}$ ions, while the electrons are mainly trapped on 
excited levels by such newly created double ionized thallium atoms or by 
previously existing ones due to thermal excitation. They form $(Tl^+)^*$ 
excited ions which decay afterwards by emitting the scintillation of energy 
$h\nu$, in the characteristic 550~nm~ band; here $h$ is the Planck constant 
and $\nu$ is the light frequency. The scheme of the process is:
\begin{eqnarray}
	Tl^{++} + electron &\longrightarrow& (Tl^+)^* + h\nu 
\nonumber		
	\\
	Tl^+    + hole &\longrightarrow& Tl^{++} .
\nonumber
\end{eqnarray}
The mathematical formalism is the same if a more complex configuration of the
luminescence centre is considered. During the non-equilibrium
state in the cylinder around the ionizing particle trajectory, the two states 
of thallium feed one another via the carrier trapping, closing in this way 
the cycle which insures the restoration of the equilibrium and the local 
electrical neutrality. A time dependent formalism which treats the kinetics of 
the volume concentrations for both types of carriers and thallium traps would
somehow cover also the ``exciton'' scenario 
\cite{mey61,bir64,smo79,spa94,nag94,nag95}. In this way, it would not 
be necessary to isolate in time the two steps of the scintillation process:
prompt formation of exciton and its delayed deexcitation.   
\par
Similarly to the levels created by thallium impurities, deeper levels
may be intruded by other point defects of the crystal: the cations and anions
removed towards interstitial positions, or for the pure alkali halides more
probably to the crystal surface \cite{kit66}, and the associated vacancies.
The positive and negative defects are produced by thermal vibrations in
practically equal concentrations $N_D$ \cite{kit66}. These levels may compete
with the activator sites in carrier trapping, following radiative transitions
in the blue band or non-radiative transitions.
In insulators, the Fermi-Dirac statistics of the carrier states is well
approximated by a Boltzmann function. The concentration of those carriers
located at the defect sites, together with the neutrality condition \cite{con70},
provide the defect distribution, described by a Boltzmann function too.
 The energy of a nearest-neighbour bound in a solid is of the order of 1 eV.
Actually, after Kittel \cite{kit66}, the cation and anion defect
concentrations are $N_D\approx N_H\times \exp(-W_D/k{\cal T})$, where
$N_H$ is the host volume concentration of cations and anions and
$W_D\approx 1$eV is much larger than the mean vibrational energy.
Comparable values for $W_D$ (0.2 - 0.4 eV) are obtained with the above
expression from concentration of elementary colour centres 
of the order of 10$^{15}$ - 10$^{18}$ centres~cm$^{-3}$, determined by means 
of absorption coefficients \cite{mar79}. In a Boltzmann type distribution,
$N_D$ would represent the number of cations or anions per unit volume having
energies higher than $W_D$. Consequently, $W_D\approx$ 1 eV would be the
minimum energy, transferred to a lattice ion, required to produce a defect.
At a given temperature, the number of defects may be 
enhanced by the nuclear interaction of the incident particle 
with the lattice nuclei. This interaction is proportional to the
specific nuclear stopping power $-(\d E/\d x)_n=S_n$ and produces an additional
defect concentration (practically the same for cations and anions) $N_n$.
 \par
 Besides, one has to consider the capture of electrons by the ionized atoms
 in the wake of the particles, i.e. the ``direct'' electron-hole
 recombination, leading to the crystal intrinsic transitions in the UV band.
 This process probability increases with the specific electronic stopping power
 $-(\d E/\d x)_e=S_e$ of
 the incident particle, while it has to be practically null for
 electrons. In experiments, a small UV peak was observed for the
 heaviest ions of low energy. On the contrary, for very energetic protons
 punching out through a thin CsI(Tl) crystal, in such a way that the Bragg
 peak is not contributing to the scintillation, both quantities $S_e$ and $S_n$
are so weak that the shape of their signal is 
 nearly similar to that of electrons generated by $\gamma$ -- rays.
 We do not consider the possible feeding of the yellow band by
 scintillations in UV band, a two step process in CsI(Tl) put in evidence by
 UV laser irradiation \cite{pou95}. 
\par
By quantifying these assumptions, it will be possible to account for the
 main experimental observations mentioned by items a) -- c) of the 
 subsection \ref{subsect31}.
\subsection{Stopping powers\label{subsect33}}
\begin{figure}
\epsfxsize=12.cm
\epsfysize=12.cm
\epsfbox{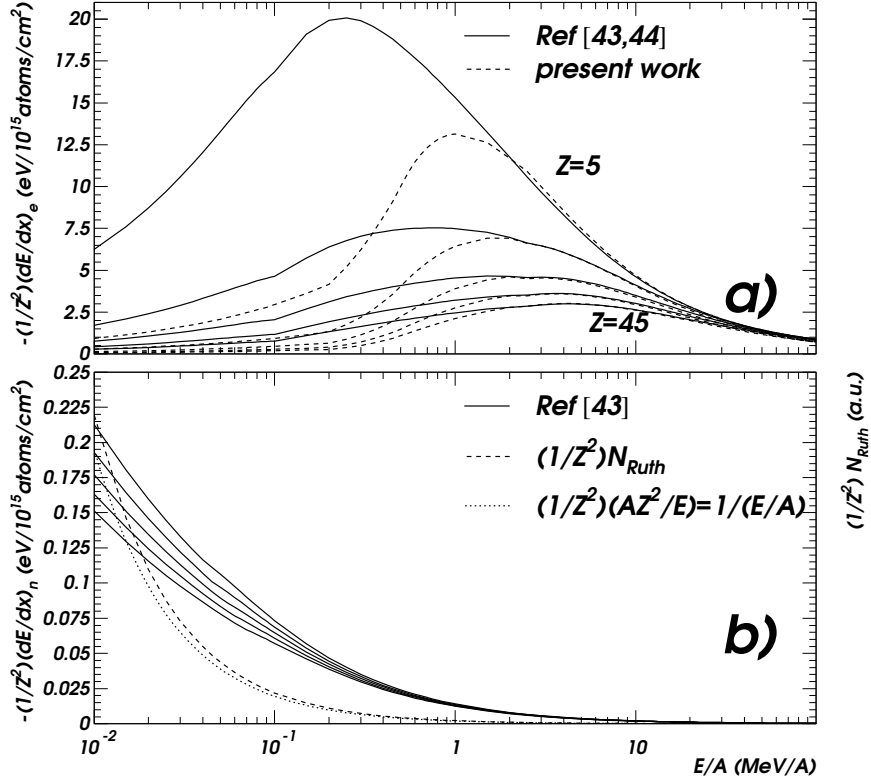}
\caption{Specific energy loss divided by $Z^2$ for different ions ($\Delta Z= 10$) in
	CsI(Tl), versus the energy per nucleon. a) Specific electronic stopping power 
	as in Refs. \cite{zie77,hub90} - solid lines; approximate specific 
	electronic stopping power 
	used in calculation of $\cal F$ - dashed lines. b) Specific nuclear 
	stopping power as in Ref. \cite{zie77} - solid lines. The dashed curves 
	concerning the density of lattice ions displaced by Coulomb scattering 
	(Eq. (\ref{ecu10}) 
	for $W_D = 1 eV$) are indistinguishable for all ions, and well 
	approximated by the first term of the mentioned formula - dotted line. 
	For a) and b), see subsection \ref{subsect33} for explanations.}
\label{fig4}
\end{figure}

An incident fragment slowing in a CsI(Tl) crystal loses its energy mostly by
ionization, and in a much smaller fraction, by interacting
with the nuclei of the lattice. The specific
energy loss is $-(\d E/\d x) = S_e + S_n = S_e (1 + S_n/S_e)$,  
 the sum of the specific electronic and nuclear stopping powers, 
expressed in eV$\times 10^{-15}$~atoms$^{-1}$~cm$^2$. 
 Multiplied by the crystal density $\rho$, they lead to the corresponding
 stopping powers per unit length: $\rho \times S_{\mathrm e,n}$, expressed in
 MeVcm$^{-1}$.
 For the calculation of the differential light output, connected to the
 differential deposited energy, we have used the stopping powers of Ziegler
 \cite{zie77}, derived from master curves of $\alpha$ particles, based on
 Bethe-Bloch formula. For the specific electronic stopping power
 $S_e$ at energies above 2.5 AMeV, an effective charge
 $Z_{eff}=Z\times\gamma_{eff}(E,A,Z)$ has been used with the parameterization
 of Hubert et al. for $\gamma_{eff}(E,A,Z)$ \cite{hub90}. Below 2.5 AMeV, the
 parameterization of Ziegler was chosen for $\gamma_{eff}(E,A,Z)$
 renormalized at 2.5 AMeV, but not lower than 200 AkeV, where
 $\gamma_{eff}(E,A,Z)$ was kept constant, at the value reached at 200 AkeV:
 $\gamma_{eff}(E=200$ $A{\mathrm keV},A,Z)$, for each type of fragment. At
 the atomic number $Z$, the integer of
 $A = 2.072Z + 2.32\times 10^{-3}Z^2$ for $Z < 50$ or
 $A = 2.045Z + 3.57\times 10^{-3}Z^2$ for $Z\geq 50$ \cite{cha98}
is used as mass number \cite{par00ii}. The resulting $S_e$ curves vs the
energy per nucleon $E/A$ are plotted - with solid lines - in Fig. \ref{fig4}a), 
for several atomic numbers. The corresponding specific nuclear stopping 
powers, as calculated by Ziegler's approach \cite{zie77}, are shown in 
Fig. \ref{fig4}b) - solid lines. 
\par
In order to determine ${\cal F}(E)$ in the next subsection, we introduce, 
besides the specific electronic stopping power $(-\d E/\d x)_e$, the 
restricted mean rate of specific electronic energy loss 
$(-\d E/\d x)_{T<T_{\mathrm cut}}$ \cite{agu94}, which is the mean rate of 
energy deposited for collisions excluding energy transfers $T$ to electrons 
greater than a cut off value $T_{\mathrm cut}$. In this case, we have preferred
analytical expressions of these quantities instead of the above mentioned
Ziegler recipe which is used later for  calculating light output and which is
not easily to handle. These are based on Bethe-Bloch expressions that are
valuable for moderately relativistic charged particles ($E/A \geq 80 MeV/u$
for the CsI medium), other than electrons \cite{agu94}, with the
parametrization $I=16Z_{\mathrm M}^{0.9}$ for the
ionization ``constant'' \cite{agu90}. By grafting on these expressions the effective charge defined above 
$Z_{eff}=Z\gamma_{eff}$, the lower limit of their domain of applicability 
goes down towards 1 - 2 MeV/u, as shown by the dashed curves in 
Fig. \ref{fig4}a) for $(-\d E/\d x)_e$.
\par
The elastic Coulomb scattering on the crystal lattice nuclei - the main part
of the nuclear interaction and hence of the specific  nuclear stopping power -
is used to quantify the density of lattice defects induced by the incident 
particle:  
$N_{\mathrm Ruth} = (\rho{\cal N}_{\mathrm A}/
(\langle A_{\mathrm CsI} \rangle
\times \pi r_c^2)) \int_{\theta^\prime_{\mathrm
min}}^\pi(d\sigma^\prime/d\Omega^\prime)d\Omega^\prime$.
Here $\langle A_{\mathrm CsI} \rangle=129.905$ g is the mean atom-gram,
$\cal N_A$ is Avogadro's number 
and (${\mathrm d}\sigma^\prime/{\mathrm d}\Omega^\prime$) is the Rutherford 
cross section in the centre of mass (c.m.) frame; the minimum deflection angle 
in the c.m. frame $\theta^\prime_{min}$, necessary to transfer an energy 
$\delta E = W_D$ in one Coulomb deflection event, may be calculated in 
classical kinematics from the latter equality if the energy required to 
produce a defect in the lattice $W_D$ is known. Then:
\begin{equation}
N_{\mathrm Ruth}\propto \left(\frac{\langle Z_{CsI} \rangle}{\langle A_{CsI}
\rangle}\right)^2
\frac{AZ^2}{E}\left[\frac{2}{W_D}-
\frac{\langle A_{CsI} \rangle}{2AE}\left(1+\frac{A}{\langle A_{CsI}
\rangle}\right)^2\right] ,
\label{ecu10}
\end{equation}
where $\langle Z_{CsI}\rangle=54$ is the average atomic number of the CsI
medium. $N_{\mathrm Ruth}$, the simple, analytical alternative to the volume 
concentration $N_n\propto S_n$, is plotted 
in Fig. \ref{fig4}b) for $W_D=1 eV$; the different atomic number
curves are practically superposed - dashed line.
In the same figure, the first term of the above equation $\propto AZ^2/E$,
weighted also by $1/Z^2$, was represented with a dotted line. This result
provides a good approximation to $N_{\mathrm Ruth}$.
\subsection{Calculation of ${\cal F}$\label{subsect34}}
The discrete functions in Fig. \ref{fig3} can not be explained in terms of the 
carrier diffusion process only. Let us suppose that there is a carrier
concentration threshold, below which quenching does not exist. The
solution of a diffusion equation in cylindrical geometry predicts a
concentration which diminishes with time and with the radial distance,
but remains proportional to the initial density of carriers per
unit length of the ion track. This fact may be only compatible with a
monotonic decrease of the light output with the stopping power and not with
the discrete functions mentioned above. In addition, the transverse outflow of 
the carriers is somehow inhibited by space charge effects. A very high 
carrier concentration is maintained in the proximity of the incident 
particle path \cite{sei73}. Consequently, the diffusion is not
taken into consideration in the next calculation.
On the other hand, we have seen in subsection 
\ref{subsect62} that this dependence is compatible with the knock-on
electron generation. These electrons have ranges greater than $r_c$, viz. 
energies higher than a cut off value $T_{cut}$, taking with them a fraction 
${\cal F}(E)$ of the energy deposited in the path element $\d x$.
\par
\begin{figure}
\epsfxsize=12.cm
\epsfysize=12.cm
\epsfbox{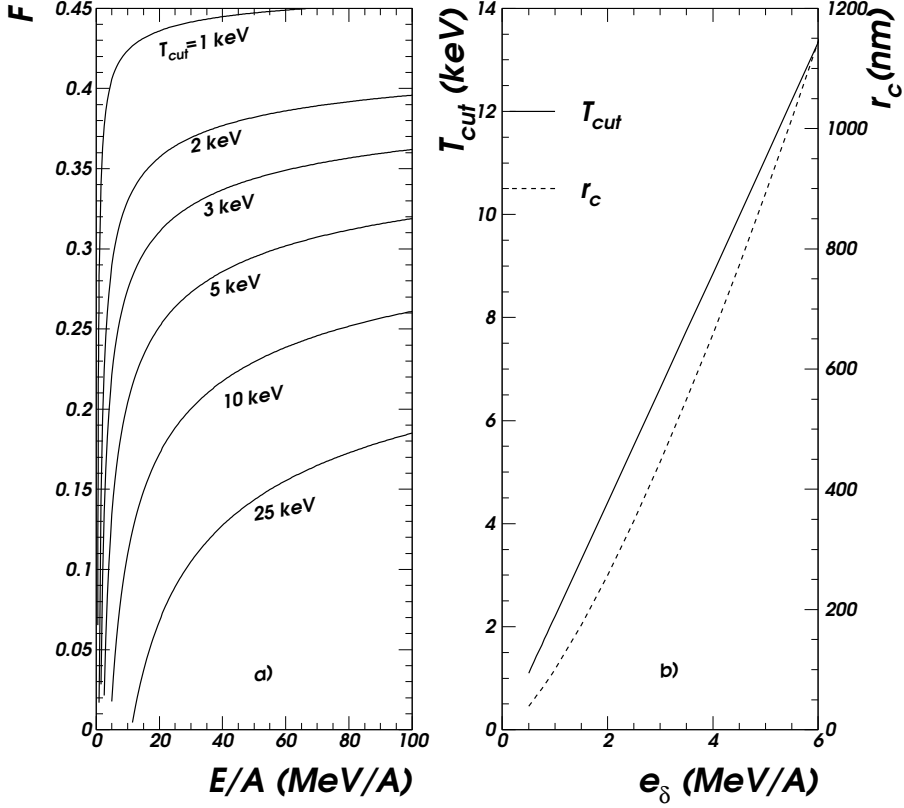}
\caption{a) The fractional energy loss $\cal F$ carried outside the primary
	column by the $\delta$ -- rays; it depends on the fragment velocity
	$\propto \sqrt{E/A}$,
	but not on its identity (Eq. (\ref{ecu15})). Different curves
	correspond to different hypothetical values of $T_{cut}$, the cut off
	value in the electron energy above which they may be considered as
	$\delta$ - rays escaping the highly ionized fiducial cylinder of 
	radius $r_c$ along the fragment wake. b) $T_{cut}$ - solid line - and 
	$r_c$ - dashed line - against the fragment energy per nucleon 
	threshold necessary to generate $\delta$ -- rays. }
\label{fig5}
\end{figure}

If $\beta$ and $\gamma$ are the kinematic variables of the incident particle,  
$c$  is the speed of light in vacuum and $u$ the unified atomic mass unit, the 
maximum energy $T_{max}$ which may be transferred to a stationary unbound 
electron of mass $m_{\mathrm e}$ may be calculated as 
$T_{\mathrm max}=2m_{\mathrm e}c^2\beta^2\gamma^2$
in the ``low energy'' approximation 
$2\gamma m_{\mathrm e}/(A\times u)\ll 1$  \cite{agu94}.
The condition that such an electron escapes the primary column and can be 
considered as a $\delta$ - ray is that this energy is at  least equal to 
$T_{cut}$; otherwise said:
\begin{equation}
 T_{max}\geq T_{cut} = 2m_{\mathrm e}c^2\beta_{\delta}^2\gamma_{\delta}^2 ,
 \label{ecu12}
\end{equation}
 where the index $\delta$ at the kinematic variables refers to the minimum energy per nucleon $e_\delta$
 of the incident particle that still generates $\delta$ -- rays
 ($\beta_{\delta}^2\approx 2e_\delta/(uc^2)$ in the non-relativistic 
approximation). $(-\d E/\d x)_{T<T_{\mathrm cut}}$, defined in the previous 
subsection,
provides the specific electronic stopping power deposited inside the primary
column. The energy deposited by the $\delta$ - rays outside the primary column
is $(-\d E/\d x)_\delta = (-\d E/\d x)_e - (-\d E/\d x)_{T<T_{\mathrm cut}}$.
 From Eq. (\ref{ecu7}), the factor $\cal F$ 
becomes in this case:
 \begin{equation}
 {\cal F} = \frac{(-\d E/\d x)_\delta}{(-\d E/\d x)_e} = 
 \frac{1}{2}\frac{\ln(\frac{T_{\mathrm max}}{T_{\mathrm cut}})-\beta^2+
 \beta^2\frac{T_{\mathrm cut}}{T_{\mathrm max}}}
 {\frac{1}{2}\ln(\frac{2m_{\mathrm e}c^2\beta^2 \gamma^2}{I^2}T_{\mathrm max})
 -\beta^2-\frac{\Delta}{2}} ,
 \label{ecu14}
 \end{equation}
($\Delta$ being the density effect in Bethe-Bloch formula
\cite{agu94}),
which in the nonrelativistic approximation $\beta^2\ll 1$ reads:
 \begin{equation}
 {\cal F} \approx \frac{1}{2}\frac{\ln(\frac{\beta^2}
 {\beta_{\mathrm \delta}^2})}{\ln(\frac{2m_{\mathrm e}c^2}{I}
 \beta_{\mathrm \delta}^2) + \ln(\frac{\beta^2}
 {\beta_{\mathrm \delta}^2})} .
 \label{ecu15}
 \end{equation}
Eq. \ref{ecu15} is valuable only for $\beta\geq \beta_\delta$, under this
limit $\cal F$ is zero. The only unknown quantity in the expression of 
$\cal F$ is $e_\delta$, which will be one of the free parameters of the model. 
The factor $\cal F$ vs $E/A$ is plotted in Fig. \ref{fig5}a) for different 
threshold energy per nucleon $e_{\mathrm \delta}$ values.
\subsection{Estimation of $r_{\mathrm c}$\label{subsect35}}
One may calculate, from equation (\ref{ecu12}), the minimum $\delta$ -- ray
energy values $T_{\mathrm cut}$ required to escape the primary column, as a 
function of $e_{\mathrm \delta}$. In the knock-on electron energy interval 
of interest here, it is found experimentally that the practical \cite{mey62} 
range - energy relation can be reasonably described in various stopping media 
by a function of the form: $R_{\mathrm e}=bT^n$ \cite{whi12}, where both b and 
n are constants \cite{kan62,kan61}. With $R_{\mathrm e}$ in mg~cm$^{-2}$ and
 $T$ in keV, and considering CsI as having an intermediate mean 
 atomic number between those of aluminium and gold, the two extreme media
 treated in Refs. \cite{kan62,kan61}, we find the particular value b = 0.016 by 
 keeping n=1.35 \cite{mey62,kan62} as universal power index for all the 
 stopping media. Consequently, and without taking into account the
 angular distribution of the scattered $\delta$ -- rays, one may estimate the
 corresponding primary column radius $r_{\mathrm c}$. Both quantities:
 $T_{cut}$ and $r_c$  are plotted in Fig. \ref{fig5}b) versus $e_{\delta}$.
\subsection{The formalism\label{subsect36}}
Inside the primary column, the initial (t=0) volume concentrations of
the created pair carriers: electrons (e) and holes (h), at the coordinate $x$
 along the particle path, are proportional to the electronic stopping power
 per unit length $\rho S_e$: 
\begin{eqnarray}
\label{ecu16}
n_{\mathrm h}(x,0)=n_{\mathrm h_0}(x) & = & 
\frac{\rho S_{\mathrm e}(x)}{\epsilon\pi r_{\mathrm c}^2}=n_0(x) \\
n_{\mathrm e}(x,0) = n_{\mathrm e_{\mathrm 0}}(x) =\left(1-{\cal F}(x)
\right) \frac{\rho S_{\mathrm e}(x)}{\epsilon\pi r_{\mathrm c}^2} & = & 
 \left(1-{\cal F}(x)\right) n_0(x) 
\label{ecu17}
\end{eqnarray}
for holes and electrons, respectively. The number of knock-on electrons
per unit length escaping out of the primary column (as long as
$E/A > e_\delta$) 
is ${\cal F} \rho S_{\mathrm e}/\epsilon$ ($\epsilon$ being the mean ionization
energy). They will produce other ionizations outside the primary column.
The radial diffusion is not treated here. 
 \par
 The basic assumptions concerning the thallium activator centres in the
 caesium iodide crystal is that they may exist in two different states: one
 acting as hole trap - of concentration $\Delta N_A$ - and the other one -
 of concentration $N_A$ - which acts as electron trap, to produce afterwards
 the scintillation.
 \par
 The excited system inside the primary column has the tendency to come back to
 the equilibrium state. The electrons 
may be trapped by thallium electron traps with a probability $\Laena$, 
 or by defects with a probability $\Ldendnn$, or they may directly recombine 
 with holes at ionized cation and anion sites of the lattice, with a 
 probability $\LRnh$. The holes, in their turn, may be trapped by the 
 thallium hole traps with a probability $\Lahna$, or by defects, with a
 probability $\Ldhndnn$ or they may recombine with the electrons, with a
 probability $\LRne$. The equations concerning the
 time variation of the carrier concentrations inside the primary column read:
 \begin{eqnarray}
 -\frac{\d n_{\mathrm e}(x,t)}{\d t} & = & \left[\Laena +
 \Ldendnn \right] n_{\mathrm e}(x,t) \nonumber \\  
 & + & \LRnh n_{\mathrm e}(x,t) 
\label{ecu18}
\end{eqnarray}
 \begin{eqnarray}
-\frac{\d n_{\mathrm h}(x,t)}{\d t} & = &
\left[\Lahna + \Ldhndnn \right] n_{\mathrm h}(x,t) \nonumber \\
& + & \LRne n_{\mathrm h}(x,t) . 
\label{ecu19}
\end{eqnarray}
By disregarding the activator centre depletion \cite{gwi63i}, the
coefficients $\Laena$ and $\Lahna$ may be considered as being
approximately constant. For the sake of simplicity, we make the same
hypothesis for $\Ldendnn$ and $\Ldhndnn$. In this way, we must solve two
coupled first order differential equations with constant coefficients. The 
initial conditions are given by the expressions
(\ref{ecu16},\ref{ecu17}), where ${\cal F}> 0$ for $E/A > e_{\mathrm \delta}$
and ${\cal F}=0$ for $E/A\leq e_{\mathrm \delta}$. Of course the
number of defects is much smaller than that of the activator centres, but
we shall see in the next section that the nuclear induced defects play a role 
in describing the total light output at lower incident energies.
In order to keep the number of free parameters in the resulting fit procedure
as low as possible, we make the hypotheses that:
$\Lambda_{\mathrm Dh}\approx\Lambda_{\mathrm De}=\Lambda_{\mathrm D}$.
Our tests have shown that the ratio
$(\Lahna + \Ldnd)/(\Laena + \Ldnd)\approx 1$ and that the shape of the total
 light output is not critically sensitive to its magnitude around this value. 
Thus, instead of keeping this ratio as a free fit parameter, we fix its value 
at 1. We make the notations:
 $\Lambda^\prime_{\mathrm Ae,D,R}=\Lambda_{\mathrm Ae,D,R}/(\Laena+\Ldnd)$.
 Under the above assumptions, one may calculate the first derivative of the 
 hole concentration with respect to the electron concentration:
 \begin{equation}
 \frac{\d n_{\mathrm h}\left(x,t\right)}{\d n_{\mathrm e}\left(x,t\right)}=
 \frac{\left(1+\Lambda^{\prime}_{\mathrm D}N_{\mathrm n}
 +\Lambda^{\prime}_{\mathrm R}n_{\mathrm e}\left(x,t\right)\right)n_{\mathrm
 h}\left(x,t\right)}
 {\left(1+\Lambda^{\prime}_{\mathrm D}N_{\mathrm n}
  +\Lambda^{\prime}_{\mathrm R}n_{\mathrm h}\left(x,t\right)\right)n_{\mathrm
  e}\left(x,t\right)} .
 \label{ecu20}
 \end{equation}
 The carrier concentrations are correlated. At every moment $t$ and position
 $x$, the hole concentration $n_{\mathrm h}\left(x,t\right)$ may be considered
 as a function of the electron concentration $n_{\mathrm e}\left(x,t\right)$,
 expressed by a Taylor series expansion around the initial value at the $x$
 coordinate $n_{\mathrm e}\left(x,0\right)=n_{\mathrm e_{\mathrm 0}}$. By
 keeping only the zero and first order terms in the expansion, viz. by
 considering that $n_{\mathrm h}\left(x,t\right)$ is linearly dependent on
 $n_{\mathrm e}\left(x,t\right)$ and by replacing it in the rate equation for 
 the electron concentration Eq. (\ref{ecu18}), one may determine the time 
 dependent solution:
 \begin{eqnarray}
\lefteqn{ n_{\mathrm e}\left(x,t\right)  =}  \\ \nonumber 
  & & \frac{p\left(x\right)\left(1-{\cal F}\left(x\right)\right)n_{\mathrm 0}
 \left(x\right)}{\left(1+\Lambda^{\prime}_{\mathrm D}N_{\mathrm n}
 \left(x\right)
 +\Lambda^{\prime}_{\mathrm R}n_{\mathrm
 0}\left(x\right)\right)\exp\left(p\left(x\right)\left(\Laena +
 \Ldnd\right)t\right) - y\left(x\right)\Lambda^{\prime}_{\mathrm R}n_{\mathrm
 0}\left(x\right)}  
 \label{ecu21}
 \end{eqnarray}
 with $p(x)=1+\Lambda^\prime_{\mathrm D}N_{\mathrm n}(x)
  +\left(1-y(x)\right)\Lambda^{\prime}_{\mathrm R}n_{\mathrm 0}(x)$
in every point $x$ of the ionizing particle trajectory, and
$y(x)=1-{\cal F}\left(x\right)\Lambda_{\mathrm R}^{\prime}
n_{\mathrm 0}(x)/\left(1+\Lambda_{\mathrm D}^{\prime} N_{\mathrm n}(x) +
\Lambda_{\mathrm R}^{\prime} n_{\mathrm 0}(x)\right)$.  
\par
For the $\delta$ -- rays, which are supposed to produce new ionizations 
outside the primary column, the new created carrier concentration equations 
are completely decoupled for electrons and holes, because there is no
radiation damage nor are there recombination processes. In the electron 
equation of interest, only the first two linear terms (concerning the 
activator and permanent defect sites) on the right side of the 
Eq. (\ref{ecu18}) have to be kept. Thus, the solution will be a simple 
exponential.
\subsection{The light output\label{subsect37}}
The double differential light output $(\d^2L(x,t)/\d x \d t)_p$ in the primary
column is given, within a multiplicative constant ``a'' (related to the energy
- light unit conversion, the light collection in connection to the shape of
the crystal and the PMT gain), by the first of the linear terms in equation
(\ref{ecu18}): $a \epsilon \pi r_c^2\Laena n_e(x,t)$.    
Integrated over time between $0$ and $\infty$, this term leads to the 
differential light output in the infinitesimal element $\d x$ of the primary 
column:
\begin{equation}
\left(\frac{\d L(x)}{\d x}\right)_{\mathrm p}=
a \epsilon \pi r_{\mathrm c}^2 \Lambda_{\mathrm Ae}^\prime N_{\mathrm A}
\frac{1-{\cal F}(x)}{y(x) \Lambda_{\mathrm R}^\prime}
\ln\left[\frac{1+\Lambda_{\mathrm D}^\prime N_{\mathrm n}(x) +
\Lambda_{\mathrm R}^\prime n_{\mathrm 0}(x)}
{1+\Lambda_{\mathrm D}^\prime N_{\mathrm n}(x) +
\left(1-y(x)\right)\Lambda_{\mathrm R}^\prime n_{\mathrm 0}(x)}\right] .
\label{ecu22}
\end{equation}
The differential light output generated by the $\delta$ -- rays in the same
infinitesimal element $\d x$ (determined by integrating over time the 
corresponding exponential solution of electron rate equation) is:
\begin{equation}
\left(\frac{\d L(x)}{\d x}\right)_{\mathrm \delta}=
a \Lambda_{\mathrm Ae}^\prime N_{\mathrm A}{\cal F}(x) \rho S_{\mathrm e}=
a \epsilon \pi r_{\mathrm c}^2 \Lambda_{\mathrm Ae}^\prime N_{\mathrm A}
{\cal F}(x) n_{\mathrm 0}(x) .
\label{ecu23}
\end{equation}
The total differential light output $(\d L(x)/\d x)$ is the sum of the light 
produced inside the primary column and outside it. 
The integration over the range of the particle $R(E_{\mathrm 0})$ or, by
changing the variable, over the energy $E_0$ leads to the total
light output expression:
\begin{eqnarray}
\nonumber
L & = & a_{\mathrm G}\left[\int^{E_\delta}_0\frac{1}{a_{\mathrm R}S_{\mathrm
e}(E)}
\ln \left(1-\frac{a_{\mathrm R}S_{\mathrm e}(E)}{1+a_{\mathrm n}S_{\mathrm
n}(E)+
a_{\mathrm R}S_{\mathrm e}(E)}\right)^{-1}\frac{\d E}{1+\frac{S_{\mathrm
n}(E)}{S_{\mathrm e}(E)}}\right. \\
\nonumber
& + &
\int_{E_\delta}^{E_0}\frac{1-{\cal F}(E)}{y(E)a_{\mathrm R}S_{\mathrm e}(E)}
\ln \left(1-\frac{y(E)a_{\mathrm R}S_{\mathrm e}(E)}{1+a_{\mathrm n}S_{\mathrm 
n}(E)+a_{\mathrm R}S_{\mathrm e}(E)}\right)^{-1}\frac{\d E}{1+\frac{S_{\mathrm
n}(E)}{S_{\mathrm e}(E)}} \\
& + & 
\left.\int_{E_\delta}^{E_0}\frac{{\cal F}(E)\d E}{1+\frac{S_{\mathrm n}(E)}
{S_{\mathrm e}(E)}}\right] ,
\label{ecu24}
\end{eqnarray}
where $y(E)=1-{\cal F}(E)a_{\mathrm R}S_{\mathrm e}(E)/(1+a_{\mathrm
n}S_{\mathrm n}(E)+a_{\mathrm R}S_{\mathrm e}(E))$.
This light output expression contains four model parameters: 
 $a_{\mathrm G}=a \Lambda_{\mathrm Ae}^\prime N_{\mathrm A} \rho$,
 $a_{\mathrm R}=\Lambda_{\mathrm R}^\prime \rho/
\left(\epsilon \pi r_{\mathrm c}^2\right)$,
 $a_{\mathrm n}=\Lambda_{\mathrm D}^\prime \rho/
\left(\epsilon \pi r_{\mathrm c}^2\right)
\times \left(\epsilon/\epsilon_{\mathrm n}\right)$,
written 
like this because 
$N_{\mathrm n}(E) = \rho S_{\mathrm n}(E)/\left(\epsilon_{\mathrm n} \pi r_{\mathrm c}^2\right)$
 (here, $\epsilon_{\mathrm n}$ is the mean energy required to produce a
dislocation) and $e_{\delta}$ ($E_{\delta}=A\times e_{\delta}$). 
These parameters denote meaningful quantities. $a_{\mathrm G}$ is the ``gain'' 
parameter, related to the energy-light unit conversion, light collection, 
conversion factor for producing photoelectrons and the electronic gain.
$a_{\mathrm R}$, $a_{\mathrm n}$ are the ``recombination quenching'' and
``nuclear quenching'' parameters respectively, related to the processes which 
divert part of the deposited energy - what we call ``quenching''. 
$e_{\delta}$ is the energy per nucleon above which the $\delta$ -- rays play 
a role in the scintillation mechanism. 
\par
The first term in Eq. (\ref{ecu24}) refers to the last part of the
particle range, where the energy has decreased below the $\delta$ -- ray
production threshold $E\leq A\times e_{\delta}$ and thus ${\cal F}(E)=0$.
The last two terms concern the first part of the trajectory 
($E>A\times e_{\delta}$ and ${\cal F}(E)>0$) as follows: the second term 
reports on the light output originating inside the primary column, while the 
third one is connected to the light produced by the knock-on electrons 
outside the primary column. For the low incident energies 
$E_{\mathrm 0}\leq A\times e_{\delta}$, ${\cal F}(E)=0$ along the entire range 
of the particle and only the first term will be present.
\section{Comparison to the experimental data\label{sect4}}
\subsection{The light output -- energy relation\label{subsect41}}
It is possible now to confront the light output - energy relation in
Eq. (\ref{ecu24}) to experimental dependence, provided that the above four
coefficients are known. Actually, they are determined, from data concerning 
ions with $Z \leq 45$, as fit parameters by a $\chi^2$ minimization procedure 
using the MINUIT package from CERN library. The experimental light output 
values $Q_{\mathrm 0}$, and the corresponding energy $E_{\mathrm 0}$ deposited in a CsI(Tl) crystal of INDRA
(see the accompanying paper \cite{par00ii}) are used for the comparison.
The integral in expression (\ref{ecu24}) is performed numerically. The
resulting fit parameter values are given in Table \ref{table1}a). The fourth
parameter $a_{\mathrm n}$ was held constant
after a preliminary analysis. Using these values in expression (\ref{ecu24}),
the total light output was recalculated and compared to the experimental
$Q_0$ values; the results are plotted versus $E_0$ in Fig. \ref{fig6} - solid
lines. There is excellent agreement, within 3\% for most data, as will be shown
later. In the inset of figure, the low energy results are shown and compared
to the case where the nuclear quenching term  $a_{\mathrm n}S_{\mathrm n}(E)$
is disregarded (dashed line). The small unrealistic bump of this pure
recombination quenching case, induced by the Bragg peak in the specific
electronic stopping power $S_{\mathrm e}(E)$, is completely eliminated by
including $a_{\mathrm n}S_{\mathrm n}(E)$. Thus, the nuclear
quenching term may be considered as a ``fine tuning'' device allowing a more 
accurate description of the experimental light output at the lowest energies.
\par
Finally, when fixing $a_{\mathrm n}$ at the proposed value
(3.9$\times 10^{-1}$ eV$^{-1}10^{15}$atoms$^{-1}$cm$^2$), it is very
important to note that light ions up to $Z = 7$ are necessary but sufficient
to determine the gain parameter $a_{\mathrm G}$ to within 10\% and the
$a_{\mathrm R}$ and $e_{\mathrm \delta}$ parameters to within 25\% of the above
mentioned values. If three of the parameters: $a_{\mathrm n}$,
$e_{\mathrm \delta}$ and $a_{\mathrm R}$ are fixed at the values listed in
Table \ref{table1}a), the gain parameter $a_{\mathrm G}$ is determined to
within 2.5\% when the experimental data employed in the fit procedure are 
restricted to light ions $Z \leq 7$ too.
\begin{table}
\caption{Fit parameters: $a_{\mathrm G}$, $a_{\mathrm R}$, 
	$e_{\mathrm \delta}$, $a_{\mathrm n}$: gain, recombination quenching, 
	$\delta$ -- ray production energy per nucleon threshold and nuclear
	quenching, respectively, for module 2 of ring 3. The errors on the
        parameters (one unit on the last digit) are only statistical. For 
	exact calculation (Eq. (\ref{ecu24})):
	a) $N_{\mathrm n} \propto S_{\mathrm n}$; 
	b) $N_{\mathrm n} \approx N_{\mathrm Ruth}$ and $W_{\mathrm d}$ = 1 eV,
	were assumed. 
	c) Pure recombination case. 
       	For the first order approximation calculation (Eq. (\ref{ecu26})): 
	d) $N_{\mathrm n} \propto S_{\mathrm n}$. See subsections
	\ref{subsect37}, 
	\ref{subsect45} for explanations.}
\begin{center}
\begin{tabular}{lcccc}
\hline
 & a) & b) & c) & d) \\
\hline
$a_{\mathrm G}$~[a.u.] & 20.00 &  19.94 & 20.57 & 18.74 \\
$a_{\mathrm R}$~[eV$^{-1}10^{15}$atoms$^{-1}$cm$^2$]	& 2.82$\times 10^{-2}$ & 2.81$\times 10^{-2}$ &
3.04$\times 10^{-2}$ & 1.26$\times 10^{-2}$ \\
$e_{\mathrm \delta}$~[MeV$/u$] & 1.96 & 1.95 & 2.08 & 1.79 \\
$a_{\mathrm n}$~[eV$^{-1}10^{15}$atoms$^{-1}$cm$^2$]	& 3.9$\times
10^{-1}$ & 4$\times 10^{-1}$ [a.u.] &
 -   & 3.9$\times 10^{-1}$ \\
\hline
\end{tabular}
\end{center}
\label{table1}
\end{table}
\par
When the defect concentration created by nuclear collisions, instead of
being taken as $\propto S_{\mathrm n}(E)$, is 
calculated from the Rutherford diffusion probability - Eq. (\ref{ecu10}) - 
the fit parameter values are practically the same (eventually except for
$a_{\mathrm n}$ because $N_{\mathrm Ruth}$ may be used in arbitrary units).
These values are shown in Table \ref{table1}b) for $W_{\mathrm D}=$1 eV. The 
light output values calculated in this case are practically identical to those 
from the previous recipe. 
Thus, we conclude that the simple expression of 
$N_{\mathrm Ruth}$, very well approximated by its first term $\propto AZ^2/E$ 
as shown in Fig. \ref{fig4}b), may be used for the nuclear quenching term 
estimation. In this case, the value of the fourth parameter $a_n$ depends on 
the value used for $W_D$: the smaller $W_D$ is, the higher the number of 
created imperfections (defect or large amplitude vibration levels) and the
smaller the determined fit parameter value, and vice-versa. 
These two quantities cannot be kept simultaneously as free parameters 
in the fit procedure and one has to choose, for example, a value for $W_D$. 
\par
In the end, we have to note that the agreement of calculation 
with data is rather good (except for the lowest energies) even when the 
nuclear term is disregarded, with the advantage that the number of fit 
parameters reduces to three. The resulting parameter values are also presented
in the Table \ref{table1}c). 
\par
The quality of the above approach, which will be 
termed the recombination and nuclear quenching model (RNQM) in the following 
discussion, is illustrated also in the next subsections by three direct 
consequences.
\subsection{The $\delta$ -- ray effect\label{subsect42}}
The fraction of the deposited energy ${\cal F}(E)$ transferred to $\delta$
-- rays increases with $E/A$ - Fig. \ref{fig5}a), and has the same value 
irrespective of $Z$. For two
different fragments having the same average stopping power
$\propto AZ^2/E_{\mathrm 0}$, the higher Z fragment has a higher energy per
nucleon, a higher ${\cal F}(E)$ value and, consequently, a higher 
experimental efficiency $Q_{\mathrm 0}/E_{\mathrm 0}$, as shown by symbols in 
Fig. \ref{fig3}.
The calculated scintillation efficiency $L/E_{\mathrm 0}$ provided by RNQM
shows the same behaviour. The total light
output is given by expression (\ref{ecu24}), with the nuclear quenching term
$a_{\mathrm n}S_{\mathrm n}(E)$ or $a_{\mathrm n}N_{\mathrm Ruth}(E)$ (for
$W_{\mathrm D} =$ 1 eV - 1 keV). The results are shown by the
curves in the upper box of Fig. \ref{fig3} for a few of the heavier fragments, 
where the effect is quite pronounced. As long as the average specific 
electronic stopping power is not too high, they qualitatively follow the
experimental data. In the latter domain, the curves are increasing when the 
nuclear quenching term is neglected.
\subsection{Reaction product identification in a Si - CsI(Tl) map
\label{subsect43}}
At forward angles (rings 2 - 9), the INDRA array is composed of three detection
 layers: ionisation chamber, 300~$\mu$m thick silicon detector (Si) and 
 CsI(Tl) scintillator. The reaction products cover a large 
range in energy and atomic number and they can be well identified in a 
$\Delta E_{\mathrm Si} - Q_{\mathrm 0}$ map - symbols in Fig. \ref{fig7}  
- where $\Delta E_{\mathrm Si}$ is the energy
lost in the silicon detector. With the values of the fit parameters
$a_{\mathrm G}$,$a_{\mathrm R}$,$a_{\mathrm n}$,$e_{\mathrm \delta}$ in
column a) of Table \ref{table1} and the expression (\ref{ecu24}) for the light
output, a $\Delta E_{\mathrm Si} - L$ map was calculated - solid lines in Fig.
\ref{fig7}. The cases b), c) of Table \ref{table1} give results practically
identical to the case a). The calculated curves have excellent agreement with 
the experimental loci, within $\Delta Z \approx 1$ for $Z\approx 50$.
\begin{figure}
\epsfxsize=12.cm
\epsfysize=12.cm
\epsfbox{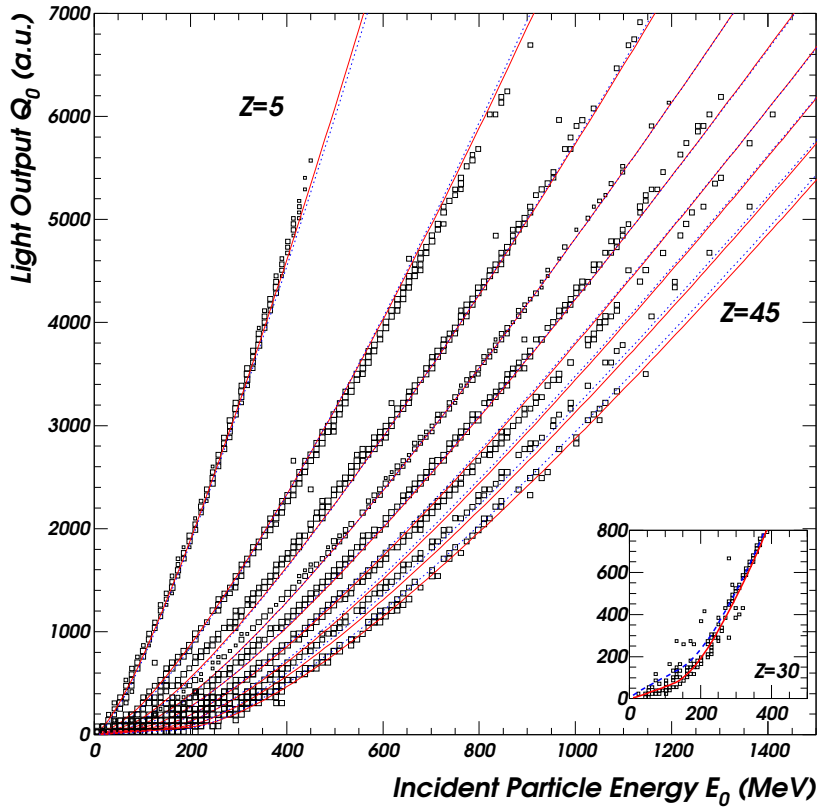}
\caption{Total light output $Q_0$ against the initial particle energy $E_0$ for
	different ions ($\Delta Z$ = 5): the symbols are experimental data from	
	 the system Xe + Sn at 32 and 50 AMeV \cite{par00ii}, the solid lines 
	are exact calculation (Eq. (\ref{ecu24})) and
	the dotted lines are first order approximation calculation (Eq.
	(\ref{ecu26})). Inset: zoom at low energies; 
	comparison to the case where the nuclear quenching term would be 
	neglected (dashed line) for $Z = 30$.}
\label{fig6}
\end{figure}
\subsection{Energy determination\label{ssubsect44}}
Having determined the four fit parameters, we may recalculate by Newton 
iterative method starting from $Q_{\mathrm 0}$, the incident energy
$E_{\mathrm calc}$ of the reaction product, to be compared to the true energy
$E_{\mathrm 0}$. For most of the data, the relative deviation of the
calculated energy versus the true energy per nucleon is less than 3\% - as
shown in Fig. \ref{fig8}a) for several ions with $5 \leq Z \leq 50$.
Only at low energies $E \leq 5$ AMeV, the discrepancy reaches rather 
large values: 5\% - 10\%, and up to 15\% in the worst cases - see Fig. 
\ref{fig8}b). Note that this roughly corresponds to the true energy accuracy, 
shown by the regions between the two solid lines in Fig. \ref{fig8}a)
\cite{par00ii}.
\begin{figure}
\epsfxsize=12.cm
\epsfysize=12.cm
\epsfbox{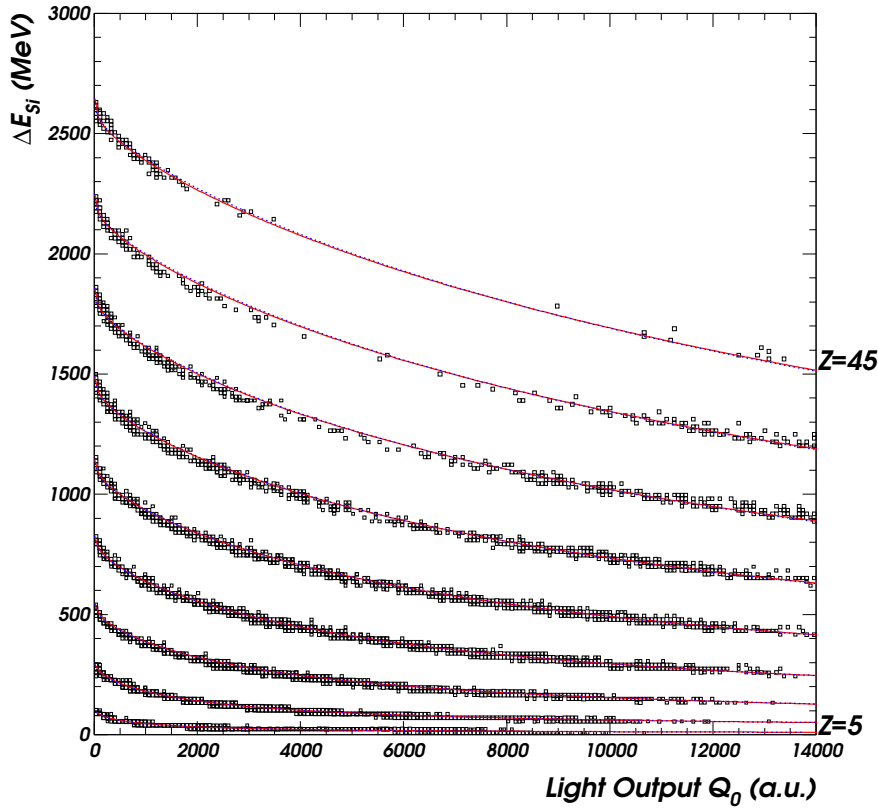}
\caption{A $\Delta E_{Si} - Q_0$ 
	map (module 2, ring 3 of INDRA) from the sytem Xe + Sn at 32 and 50 
	AMeV. 
	Conventions of presentation are the same as in 
	Fig. \ref{fig6}.}
\label{fig7}
\end{figure}
\par
These results are arguments in favour of the present ``quenching'' model
which accounts for the main experimental evidence concerning the scintillation 
yield of the CsI(Tl) crystals in the yellow band of interest.
The ratio $a_{\mathrm R}/a_{\mathrm n}$ could give an idea of the weights of
the two processes: recombination and nuclear quenching, provided that the
ratio of the mean energy required to displace a lattice ion from its site
to the mean ionization energy $\epsilon_{\mathrm n}/\epsilon$ is known.
\begin{figure}
\epsfxsize=12.cm
\epsfysize=12.cm
\epsfbox{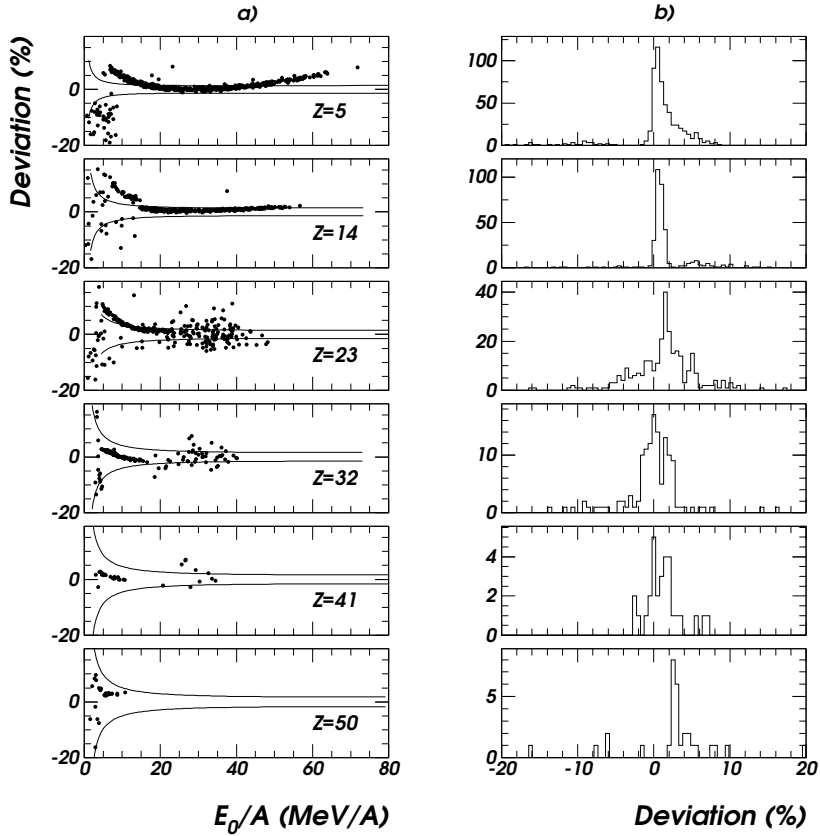}
\caption{Deviations (in \%) of the energy values determined with Eq. 
	(\ref{ecu24}) from the true energy values for several products of the 
	reactions Xe + Sn at 32 and 50 AMeV: a) deviations plotted against the 
	product energy per nucleon: symbols; 
	the regions between curves show the accuracy of the true energy per 
	nucleon (see subsection  2.2 of the accompanying 
	paper \cite{par00ii} for details); b) deviation histograms.}
\label{fig8}
\end{figure}
The parameter
$e_{\mathrm \delta} \approx 2$ MeV$/u$, found by the fit procedure, to
which corresponds a minimum energy of
4.4 keV for the $\delta$ -- rays, allows to
roughly estimate - as 
proposed in subsection \ref{subsect35} - the radius of the primary column
$r_{\mathrm c}\approx$260~nm.~ This value is one order of magnitude
greater than it has been found in Ref. \cite{mey62}. Ref. \cite{sei73}
estimates the same quantity in Si to be $1.15 \mu$m, with the connected
$\delta$ - ray energy 
$\approx 10$ keV. For electrons having this
energy, the range in CsI is of about 770~nm,~ of the same order of
magnitude as the value found in the present work. The $r_{\mathrm c}$ results 
are given in Table \ref{table2}. By means of the estimated value
$r_{\mathrm c}\approx$260~nm,~ one may determine the order of magnitude of the
carrier concentration $n_0$ generated inside the strongly ionized cylinder.
For energies 5 $\leq$ E $\leq$ 80 AMeV, one finds
$10^{15} \leq n_0 \leq 10^{18}$ carriers~cm$^{-3}$ when $2 \leq Z \leq 50$, i.e.
$n_0 \leq N_{A0}$. This fact \textit{a posteriori} supports disregarding  
activator centre depletion.   
\begin{table}
\caption{Estimation of the radius $r_{\mathrm c}$ of the transverse area of the
	primary column: a) Ref. \cite{mey62}, b) present work 
	c) Ref. \cite{sei73}.}
\begin{center}
\begin{tabular}{cccc}
\hline
 & a) Ref. $\cite{mey62}$ & b) Present work &
 c) Ref. $\cite{sei73}^*$ \\
$r_{\mathrm c}$~[nm] & 40 & 260 & 770 \\
\hline
\end{tabular} \\
\end{center}
$^*$ extracted from data in Si
\label{table2}
\end{table}
\subsection{The first order approximation of the light output
expression\label{subsect45}}
The quenching terms in Eqs. (\ref{ecu22}) 
for the differential light output and (\ref{ecu24}) for the total light output 
were evaluated by means of the parameter values from Table \ref{table1}a) and 
stopping powers in the domain of our experimental data:
$a_{\mathrm n}S_{\mathrm n} < a_{\mathrm R}S_{\mathrm e}\ll 1$. In the
first order approximation of the logarithm, the differential light
output in the primary column becomes:
\begin{equation}
\left(\frac{\d L(x)}{\d x}\right)_{\mathrm p}
\approx a_{\mathrm G}\frac{1-{\cal F}(x)}
{1+a_{\mathrm n}S_{\mathrm n}(x)+a_{\mathrm R}S_{\mathrm e}(x)}
S_{\mathrm e}(x) .
\label{ecu25}
\end{equation}
In the low energy case $E_{\mathrm 0} \leq A\times e_{\mathrm \delta}$ for 
which ${\cal F}(x) = 0$, the above expression is, except the nuclear term, 
identical with the formula of Birks \cite{bir64} from subsection
\ref{subsect61}. In the same approximation, the total light output reads:
\begin{eqnarray}
\nonumber
L & = & a_{\mathrm G}\left[\int^{E_\delta}_0\frac{1}{1+
a_{\mathrm n}S_{\mathrm n}(E) + a_{\mathrm R}S_{\mathrm e}(E)}\times\frac{\d E}
{1+S_{\mathrm n}(E)/S_{\mathrm e}(E)}\right. \\
\nonumber
& + & 
\int_{E_\delta}^{E_0}\frac{1-{\cal F}(E)}{1+
a_{\mathrm n}S_{\mathrm n}(E) + a_{\mathrm R}S_{\mathrm e}(E)}
\times\frac{\d E}{1+S_{\mathrm n}(E)/S_{\mathrm e}(E)} \\
& + &
\left.\int_{E_\delta}^{E_0}\frac{{\cal F}(E)\d E}{1+S_{\mathrm
n}(E)/S_{\mathrm e}(E)}\right] .
\label{ecu26}
\end{eqnarray}
The resulting fit parameter values are given in Table \ref{table1}d).
The quality of the results (total light output, fragment identification) 
obtained in the first order approximation is very close to that achieved by 
means of the exact treatment, as shown in Figs. \ref{fig6}, \ref{fig7} -- 
dotted lines. Formula (\ref{ecu26}), as well as expression 
(\ref{ecu10}) in the mentioned approximate form, will allow
the analytical integration of the total light output, useful for the
applications presented in section 4 of the accompanying paper \cite{par00ii}.
\section{Conclusions\label{sect5}}
The nonlinear response of CsI(Tl) was quantified in terms of the competing
deexcitation transition probabilities of the excited fiducial volume
around the path of a strongly ionizing particle. A simple mathematical
formalism was developed to determine total light output based on a
numerically integrable, four-parameter dependent expression. The formalism has
allowed the inclusion of energetic knock-on electrons. Consequently, their
effect in the scintillation efficiency was quantitatively taken into account,
in a fully consistent way, for the first time. By a fit procedure, the values
of the four parameters were determined. Of physical interest is the energy
per nucleon threshold for producing $\delta$ -- rays by the incident particle,
 $e_{\mathrm \delta}= 2$ MeV$/u$. This result also allows the rough estimation
of the primary column radius $r_{\mathrm c}$=260~nm~ along the wake of the
ionizing particle. The predictive character of the model was verified by
using telescope-type ($\Delta E_{\mathrm Si} - Q_0$)  reaction product loci 
data, 
the agreement calculation - experimental data being 
within $\Delta Z \approx 1$ for $Z\approx 50$, as well as by using deposited 
energy data, in the range 1 - 80 AMeV. A comparison of the model predictions 
to the latter experimental data resulted in variations of the order of the 
CsI(Tl) scintillator resolution $\approx 3\%$, and within 10\% -- 15\% when 
the reference energy was known with only 
a comparable, poorer precision. The effect of the 
$\delta$ -- rays in the scintillation efficiency as a function of the average 
stopping power in the entire range is quite well described too. In fact, the 
knock-on electrons play an important role in the light output, especially 
for heavy reaction products. In the accompanying paper \cite{par00ii}, 
an approximate but analytically integrable formula will be derived from the 
proposed formalism. Up to now, it was used for rapid calibration and 
identification procedures in the INDRA 4$\pi$ array. In the future, in relation
 with computer capabilities, we will use the exact formula proposed here.

\end{document}